\documentclass[12pt]{article}
\usepackage{amsmath}
\usepackage{amssymb}
\usepackage{color}
\usepackage{natbib}
\usepackage{hyperref}

\hypersetup{
    colorlinks=true,
    linkcolor=red,
    filecolor=magenta,      
    urlcolor=blue,
    bookmarks=true,
    citecolor=blue,
}

\begin{document}

\title{The Bitwise Hashing Trick \\ for Personalized Search
\footnote{
This is an original manuscript / preprint of an article published by Taylor \& Francis 
in Applied Artificial Intelligence on June 20th, 2019, available online: 
\href{http://www.tandfonline.com/10.1080/08839514.2019.1630961}{http://www.tandfonline.com/10.1080/08839514.2019.1630961} 
}}

\author{
    Braddock Gaskill\\
    eBay Inc.\\
    bgaskill@ebay.com \\
}
\date{}
\maketitle

\section{Abstract}

Many real world problems require fast and efficient lexical comparison of large
numbers of short text strings.  Search personalization is one such domain.  We
introduce the use of feature bit vectors using the hashing trick for improving
relevance in personalized search and other personalization applications.  We
present results of several lexical hashing and comparison methods.  These
methods are applied to a user's historical behavior and are used to predict
future behavior.  Using a single bit per dimension instead of floating point
results in an order of magnitude decrease in data structure size, while
preserving or even improving quality.  We use real data to simulate a search
personalization task.  A simple method for combining bit vectors demonstrates
an order of magnitude improvement in compute time on the task with only a small
decrease in accuracy.

\clearpage

\section{Overview}

\subsection{Introduction}

In personalization at eBay we take a user's chat and search history into
account to help find and rank relevant items when the user searches for
products.  Search must return results quickly; search personalization must be
fast.  eBay has more than 168 million active users; therefore personalization data
must be compact, easy, and quick to access.

Here we investigate the problem of predicting what item an eBay member will buy
based on items this user has viewed in detail in the past.  We construct an
experiment using real-world inventory and purchasing data.

Our prediction uses only the text titles of inventory items.  The essence of an
item is captured by its title and title similarity can be used to predict
sales.

eBay item titles pose both a challenge and an opportunity because sellers cram
a lot of detail in the title.  Titles are frequently not grammatically correct,
and are often incoherent.  Titles often contain strange punctuation.  Figure 1
has examples of typical eBay item titles, illuminating their peculiarity.

For title-to-title comparison, we show that a simple character n-gram vector
representation can suffice.  We further show that this representation can be
reduced to 1-bit per dimension with almost no degredation.  This technique is
simple, small, and fast for short strings.  It requires no training.

Personalized search improves user experience
\citep{Teevan2010-om,Teevan2005-mj}.  Users' ability to communicate their
``information need'' (the purpose of their search) is hampered by the imprecision
of language, homonyms, and the lack of context.  It is likely, therefore, that
adding a user's personal data, such as the text from titles of previous items
viewed, could improve search relevance and search ranking.

\begin{figure*}[t]
    \caption{Example eBay item titles}
    \textit{\small{
Sony 1-873-858-11 Video/HDMI Board, Pulled from KDL-52W3000 LCD TV *EXCELLENT* \\
GE / Hotpoint / Kenmore Oven - VENT TRIM - White - EUC! \\
1995-96 SUMMIT WAYNE GRETZKY \#24 * Los Angeles Kings HOF center \\
0574 Screw-on - Black CZ Tunnels 2 Gauge 2G Plugs 6mm \\
Adidas Yeezy Boost 350 V2 Black Core White size 9 100\% Authentic 480pp \\
Vans Classic Slip-On Damen US 6.5 Schwarz Slipper Ohne Karton 4054 \\
Authentic Genuine Original Bose IE2 MIE2i iphone remote control mic Earphones \\
}}
\end{figure*}

\subsection{Representation Size}

A common method for representing and comparing items such as
documents or listing titles is with a vector created by feature hashing aka
``the Hashing Trick'' \citep{Weinberger2009-yh,Attenberg2009-og}.  The vector is
initialized to zero.  Each feature is hashed to an index number modulo the
vector length.  The vector element at the index position is then incremented,
or in some implementations decremented based on a second sign hash.
Items are then compared pairwise using a similarity function such
as the cosine between the vectors.

In our initial approach, we found that adequate performance could be obtained
only with vectors with a minimum length of 8,000 32-bit floating point
dimensions.  Such vectors are 32 KB - too large for our scale and application.
Performance degraded significantly when dimensionality was reduced.  This
vector length problem is common in the literature \cite{bai_2009}; a few
approaches to overcoming the problem have made some progress\cite{Weinberger2009-yh}.

Our contribution is to radically reduce the type of the elements from 32-bit
floats to 1-bit bits.  This allows us to reduce the vector storage requirement
by a factor of 32, while preserving the 8,000 element dimensionality, at the
expense of not handling collisions.  The new bit arrays are 1 kbyte in size.

Grzegorczyk and Kurdziel obtained competitive results using ``Binary Paragraph
Vectors'' compared with real valued paragraph vectors in
\citep{Grzegorczyk2016-io}.  They obtained binary representations of paragraphs
from a sigmoid neural network layer, an approach that is very different from
our simple, faster hashing approach.  Hubara et al \citep{Hubara2016-ab} used
Binarized Neural Networks with weights and activations of a single bit of
precision and discovered great speed improvements with only minor performance
degradation.

\subsection{The User Vector}

In our model we represent each user using a fixed size ``User Vector'', that has
the same dimension as the item vectors.  In our task, we need to compare N
viewed items in the user's history with the M items that are search result
candidates.  We commonly call the search result candidates the ``recall set''.
We want to score each item in the recall set with the predicted relevance for
the user.

As a baseline, we could use pairwise comparisons of each of N viewed items with
each of M search candidates.  This approach would take the smallest distance
score from the N viewed items and assign it to a search result candidate.  We
use the cosine similarity to compare the vectors.  This score would be used to
re-rank the recall set.

The problem with this approach is that it takes O(M x N) comparison operations.
In our trial, the median value for N is about 44 viewed items, and the number
of items in the recall set (M) is about 100.  This approach would require 
4,400 comparisons as well as the retrieval of 44 item vectors from the user's history.

It is faster to summarize the user's view history into a single vector.
That single vector would be some combination of all N viewed items.  A single
vector would require only O(M) comparison operations to score the search
recall set of M items.

This single user history summary vector is what we call the ``User Vector''.

\section{Method}

\subsection{Character N-grams}

eBay item titles are very dense and rarely grammatically correct language.  See
Figure 1 for examples.

Instead of attempting to build a fixed word vocabulary for these noisy titles,
we chose to use character level n-grams.  We found that character 5-grams worked
well in performing lexical comparison between eBay titles.  We use 5-grams
throughout this paper as the features which get hashed into vectors.  We use
overlapping n-grams.

Using overlapping character n-grams has additional advantages when using The
Hashing Trick because it compensates for occasional collisions.  If the 3-gram
``hel'' from the word ``hello'' collided with another word's 3-gram in the array,
it is highly unlikely the 3-grams ``ell'' and ``llo'' would also collide.  This
makes words and phrases represented by character n-grams somewhat more robust
to collisions than tokenized words.

\subsection{Cosine Similarity for Bit Vectors}

Cosine Similarity is popular as a similartiy measure in the vector space model
for text retrieval\citep{Ida2008-dm}.  In vector space text retrieval, the
discrimination of syntactic elements of text is commonly used to weight each
dimension in the vector space.  Syntactic elements include words, phrases, or
overlapping N-grams.  The weights are often the output of a TF-IDF calculation
(inverse document frequency times term frequency).  

In these processes floating point vectors are commonly used.  However, as we
will see cosine similarity can be computed very efficiently for bit arrays.

Consider two real vectors

\definecolor{fgC}{rgb}{0,0,0}\color{fgC}\[\textbf{A},
\textbf{B}\in\mathbb{R}^{d}\]

the cosine similarity is defined to be:

\definecolor{fgC}{rgb}{0,0,0}\color{fgC}\[cos(\theta) = \frac{\textbf{A}\cdot \textbf{B}}{\left\Vert \textbf{A} \right\Vert\left\Vert \textbf{B} \right\Vert}
=\frac{\sum_{i=1}^{d} A_i B_i}{\sqrt{\sum_{i=1}^{d}A_i^2}\sqrt{\sum_{i=1}^{d}B_i^2}}\]

Now consider how that equation is greatly simplified when we define

\definecolor{fgC}{rgb}{0,0,0}\color{fgC}\[\textbf{A}, \textbf{B}\in\{0,1\}^{d}\]

ie, constraining each element to be a single bit which always has the value 1.0 or 0.0.  In this case, the dot
product between \textbf{A} and \textbf{B} becomes a simple boolean AND function with summation;
and the magnitude of \textbf{A} becomes the square root of the number of bits set to 1 
(population count) in \textbf{A}\citep{Hubara2016-ab}.

\definecolor{fgC}{rgb}{0,0,0}\color{fgC}\[\frac{\textbf{A}\cdot \textbf{B}}{\left\Vert \textbf{A} \right\Vert\left\Vert \textbf{B} \right\Vert} = \frac{popcnt(\textbf{A} \cap \textbf{B})}{\sqrt{popcnt(\textbf{A})popcnt(\textbf{B})}}\]

\textit{popcnt} is the ``population count'' which is defined to be the number of bits
set to 1 in the array\citep{Hubara2016-ab}.  It is implemented as a fast
hardware instruction for most modern CPU's and GPU's.

This binary cosine similarity equation is known as the \textit{Ochiai Coefficient} \citep{Ida2008-dm}.

\subsection{Vector Combination}

We use a very simple technique to construct our combined User Vector from the
item title embeddings for items previously viewed.  We simply add the individual
title vectors up element-by-element and normalize the result.  This user vector,
which is still in the same space as the individual title vectors, can then be
compared to each title vector in the recall set to score them.

For our bit vectors, we logically OR the individual title vectors into a
combined user vector.

\subsection{Data}

We obtained user activity from clickstream data of one million users over a two
week period.  This dataset includes items viewed in detail by the user (clicked
through), and clicks on a button to purchase an item (we do not know if the
checkout was fully completed).

We choose to break the user's historical activity into ``sessions'', which are
lengths of time when the user was active on the eBay site.  Among other events
such as log outs, 30 minutes of inactivity closes the current session.

We sampled 14,245 purchases from this dataset with the following
constraints:

\begin{itemize}
    \item Users were selected at random
    \item Purchase event was preceded by at least one other Session with at least
        one other viewed item
    \item Only one purchase was sampled per user
    \item if there were multiple purchases by a user, the last purchase was used
\end{itemize}

To build our simulated recall sets we leveraged category information for each
item viewed or purchased.  Here we benefit from the fact that the seller is
financially motivated to properly categorize the item for sale in one of eBay's
approximately 20,000 item categories.

\subsection{Training Free}

This technique requires no training.  There are no parameters apart from the
chosen vector dimension.  It should be insensitive to the hashing function
used as long as it is sufficiently random.  It can be used to compare short
strings where there is a lack of historical data.

\subsection{Alternative Interpretation of Bit Vectors}

A long bit vector formed by the hashing of features modulo the vector length
(``The Hashing Trick'') can be interpreted as the set of n-grams present in the
title.  We ignore collisions so duplicate occurances of n-grams are ignored -
an n-gram is either present or absent.  Because we use only one bit per
dimension we can have a far sparser vector in the same memory footprint
reducing the chance of feature collisions.

When we OR a set of item titles represented by bit vectors together to form our
combined ``User Vector'', we are essentially making a combined set of features
present in the set of titles.

\section{Experiment}

For our experiment we make a rough simulation of a search result ranking task,
without using actual search recall sets.  We take an item a user has bought,
mix it in a bag with up to 100 other random items in the same eBay item
category. Then, based on viewed items in the user's \textit{prior} sessions we
try to identify the item the user actually bought.  We score and sort all the
items in the bag and measure the accuracy for getting the bought item within
the top-1, top-5, and top-10 ranked positions.

There are a median of 44 viewed items preceding each purchase for a user
(the dataset only extends back less than two weeks).  We are experimenting with
a test set of 14,245 purchases.

The challenge of a User Vector is to collapse the embeddings from those ~44
previously viewed items into a single compact vector which can be used to
predict user behavior.  Of course viewed items is just the start, we want to
eventually include all user attributes in a vector in future work.

Using an exhaustive item-to-item match the best we have been able to do on this
task is predict with 34\% accuracy the bought item as top-1, and  44\% recall
within the top-5.  We find that somewhat remarkable because users often don't
buy something related to their activity in the prior session - it is often just
not in the data.

\begin{table*}[!t]
    \caption{Pair-wise Comparison Results}

    This table contains results for the task using comparison of each title in the
    user's history with each item title in the recall sets, using the minimum
    distance as the score.

    \begin{tabular}{l|llllll}
        Type         & dim & size(byte) & time(sec)\* & 1-top & 5-top & 10-top \\
        \hline
        \textbf{pairwise float} & 8,000 & 32,000      & 4,730s    &    \textbf{33.93\%} &    44.24\% &    50.14\% \\
        pairwise float & 1,000 &  4,000      & 1,343s    &    32.80\% &    42.40\% &    48.82\% \\
        \hline
        \textbf{pairwise 1-bit} & 8,000 &  1,000      & 1,718s    &    \textbf{33.65\%} &    44.20\% &    50.22\% \\
        pairwise 1-bit & 1,000 &    125      & 1,030s    &    32.71\% &    42.54\% &    48.70\% \\
    \end{tabular}

    \renewcommand{\arraystretch}{1.3}
    \caption{Combined User Vector Results}

    This table contains results for the task using a combined User Vector
    compared to each item title in the recall sets.

    NOTE: Python implementation was not optimized for speed.  These
    numbers give an extremely rough comparison.

    \begin{tabular}{l|llllll}
        Type          &  dim  & size(byte) & time(sec)\* &  1-top  &  5-top  &  10-top \\
        \hline
        user-vec float & 8,000 & 32,000      & 3,104s    &    29.10\% &    40.99\% &    47.74\% \\
        user-vec float & 1,000 &  4,000      &   847s    &    25.50\% &    36.81\% &    44.71\% \\
        \hline
        \textbf{user-vec 1-bit} & 8,000 &  1,000      &   253s    &    \textbf{32.83\%} &    43.90\% &    50.25\% \\
        user-vec 1-bit & 1,000 &    125      &   198s    &    19.87\% &    30.49\% &    38.23\% \\
    \end{tabular}
\end{table*}

\section{Results}

Table 1 and Table 2 contain a summary of results from our experiment.  The
columns include the dimension of the arrays, the storage size (assuming 32-bit
floats), and the execution time in seconds.  Finally, the accuracy of the
method in identifying the purchased item by a sorted ranking of the recall set
in the 1st position, top five positions, or top ten positions.  

Our best performing method (pairwise float) could predict the bought item
33.93\% of the time in the top-1 position.  ``pairwise 1-bit'', despite using a
32 times smaller data structure, trailed by only 0.28\%.  We believe from
manual inspection that that is about as good as can be achieved with this
dataset.  Many bought items are unrelated to items the user viewed in previous
sessions.

As can be seen in the results, the 1-bit vectors dominate in speed, storage,
and accuracy in the ``User Vec'' experiments.  The fact that accuracy is improved
over a much more precise floating point vector of the same dimension is 
interesting.  It appears that the fact that collisions are ignored and a vector
element can never have a value greater than 1 actually help with this dataset.
This makes sense, since our title are so dense with information that commonly
repeated n-grams which contain little information would detract from the
content of the title during the summation and normalization process.

\section{Challenges, Limitations, and Further Work}

eBay titles are relatively short strings of under 80 characters with little
repetition.  This keeps the hashed vectors sparse, which is critical when using
a simple binary OR to combine them as we are doing.  We are also only
attempting to combine a few score title vectors in this way, which keeps the
resulting combined User Vector sparse.  It is likely that this technique would
not work on long documents, or on much larger numbers of documents.  On large
numbers of long documents, the binary feature vectors are very likely to
saturate and collisions would increase.

Our sampling of user history is slightly awkward.  We followed 1 million users
who made at least one purchase over a two week period in early
November 2016.  Because a purchase may have occurred any time in the two week
window that was sampled, users who bought items early in the window would have
less history than those who bought items later in the window.  

There are a number of similarity metrics that could be applied to bit vectors.
We only experimented with the Ochiai Coefficient and the Hamming distance, but
others such as the Jaccard Coefficient may yield valid results
\citep{Jure_Leskovec_and_Anand_Rajaraman_and_Jeff_Ullman2015-gd}.

\section{Conclusions}

For many applications dimensionality is more important than precision.  When
storage space or computation speed is a priority, we found in this trial that
reducing precision to a single bit while maintaining a rich dimensionality,
greatly improved speed, storage requirements, and even accuracy. 

We also found that for short strings (eBay item titles), a simple OR'ing of bit
vectors was actually more effective for building a composite vector than
attempting to add and normalize floating point vectors.

Using both bit vector title representations and combining title representations
into a ``User Vector'' improved speed and storage size by an order of magnitude.


\bibliographystyle{chicago}
\bibliography{bibliography}

\clearpage

\end{document}